\documentclass[a4paper]{article}
\usepackage{pdfsync}
\usepackage[utf8]{inputenc}
\usepackage{graphicx}
\usepackage{amssymb,amsfonts,amsmath,amsthm}
\usepackage{hyperref}
\usepackage{graphicx}
\usepackage{multirow}
\usepackage{verbatim}
\usepackage{color}
\usepackage{bm}
\usepackage[sectionbib,round]{natbib}

\theoremstyle{plain}

\newtheorem{proposition}{Proposition}[section]
\theoremstyle{definition}

\newtheorem{remark}{Remark}[section]

\begin{document}

\title{\textbf{Multidimensional scaling of two-mode three-way asymmetric dissimilarities: finding archetypal profiles and clustering}}

\author{Mireia Mollar-Gumbau$^1$,
Aleix Alcacer$^1$,
Rafael Ben\'{\i}tez$^2$,
\\
Vicente J. Bol\'os$^2$,
Irene Epifanio$^{1,3,*}$,
\\
{\scriptsize $^1$ Dept. Mathematics, Universitat Jaume I, Campus Riu Sec, 12071, Castell\'o de la Plana, Spain.} \\
{\scriptsize $^2$ Dept. Business Mathematics, Universitat de Val\`encia,
Av. Naranjos s/n, Val\`encia, 46022, Spain.} \\
{\scriptsize $^3$ ValgrAI - Valencian Graduate School and Research Network of Artificial Intelligence.} \\
{\scriptsize e-mail\textup{: \texttt{mollarm@uji.es} (M. Mollar-Gumbau), \texttt{aalcacer@uji.es} (A. Alcacer), \texttt{rafael.suarez@uv.es} (R. Ben\'{\i}tez), }} \\
{\scriptsize \textup{ \texttt{vicente.bolos@uv.es} (V. J. Bol\'os),  \texttt{epifanio@uji.es} (I. Epifanio)}}
 \\}

\date{}

\maketitle

\begin{abstract}
Multidimensional scaling visualizes dissimilarities among objects and reduces data dimensionality. While many methods address symmetric proximity data, asymmetric and especially three-way proximity data (capturing relationships across multiple occasions) remain underexplored. The h-plot enables the analysis of asymmetric and non-reflexive relationships by embedding dissimilarities in a Euclidean space, allowing further techniques like archetypoid analysis to identify representative extreme profiles. However, no existing methods extract archetypal profiles from three-way asymmetric proximity data. This work extends the h-plot methodology to three-way proximity data under both symmetric and asymmetric, conditional and unconditional frameworks. The proposed approach offers several advantages: intuitive interpretability through a unified Euclidean representation; an explicit, eigenvector-based analytical solution free from local minima; scale invariance under linear transformations; computational efficiency for large matrices; and a straightforward goodness-of-fit evaluation. Furthermore, it enables the identification of archetypal profiles and clustering structures for three-way asymmetric proximities. Its performance is compared with existing models for multidimensional scaling and clustering, and illustrated through a financial application. All data and code are provided to facilitate reproducibility.
\end{abstract}


\section{Introduction} \label{sec0}
Multidimensional scaling (MDS) comprises a series of statistical techniques used to visualize dissimilarities between a set of objects and reduce the dimensionality. \citet{carroll1980multidimensional} proposed a taxonomy of data and defined the way and the mode. The mode indicates a set of entities, while the way represents the number of modes (with possible repetitions) that define the data array. Three-way proximity data capture relationships between pairs of entities measured across multiple occasions, such as different conditions, settings, sources, or times. They extend traditional two-way proximity data into a three-dimensional framework, where the additional dimension represents the occasion of each pairwise proximity. These data are typically organized as a series of square matrices (one per occasion) containing proximities between all object pairs, which may be symmetric or asymmetric.

A review of the literature on MDS reveals a clear imbalance: while numerous methods have been developed for symmetric proximity data \citep{delicado2025multidimensional}, considerably less attention has been given to methods addressing asymmetric proximity data, and even less to those dealing with three-way data. 

Recently, there has been a renewed interest in dimensionality reduction for asymmetric relationships, as reflected in the growing number of publications on this topic in recent years \citep{okada2024applied, olszewski2024asymmetric, olszewski2023asymmetric, di2025candecomp, bove2021methods,Urasaki_Nakagawa_Tsuchida_Tahata_2025}. A comprehensive overview of various methods for analyzing asymmetric pairwise relationships is provided by \cite{bove2018methods}. 

\cite{Epi2013} introduced a methodology based on the h-plot, which accommodates non-metric properties such as asymmetry and lack of reflexivity (i.e., positive self-dissimilarity). The core idea is to embed the variables that define proximity to or from each object, rather than embedding the objects themselves. In the h-plot, the dissimilarity matrix is treated as a data matrix comprising two variables: the dissimilarity from object $j$ to all other objects ($d_{j \cdot}$) and the dissimilarity from all other objects to object $j$ ($d_{\cdot j}$). In this representation, the Euclidean distance between two variables corresponds to the sample standard deviation of their differences, where smaller values indicate higher similarity between variables. \cite{Epi2014} employed that methodology to analyze asymmetric citation relationships between a series of journals.

Projecting the data into a Euclidean space makes it possible to apply various statistical techniques, such as archetypoid analysis (ADA) \citep{Vinue15}, as demonstrated by \cite{vinue2017archetypoid}. ADA is an unsupervised statistical learning method that identifies representative extreme cases, known as archetypoids, within a data set. These archetypoids are actual observations used to approximate all other data points as convex combinations of them. Because ADA relies on Euclidean distances and convex combinations, projecting the data into a Euclidean space, where these operations are meaningful, ensures the method’s validity. Moreover, ADA provides a human-interpretable summary of the data, as people tend to understand and interpret extremes more intuitively.

As regards the analysis of three-way asymmetric proximity data, to the best of our knowledge, no method has been proposed yet for finding archetypal profiles in such a case.

In relation to clustering, which is another unsupervised learning methodology, although there are several methodologies for one-mode two-way asymmetric data \citep{olszewski2014asymmetric,olszewski2016asymmetric}, as far as we know, only three clustering methods have been proposed for the analysis of three-way asymmetric proximity data. \cite{chaturvedi1994alternating} identified two distinct yet overlapping sets of object clusters (corresponding to the rows and columns of the data matrices and shared across all occasions), which pose considerable interpretative challenges. A more parsimonious alternative is \cite{bocci2024clustering}, whose model relies on two clustering structures for the objects, which are linked one-to-one and shared across all occasions. The first structure represents a standard partition of the objects, capturing the average levels of exchange, while the second represents an incomplete partition that models the imbalances by allowing certain objects to remain unassigned. Furthermore, to accommodate the heterogeneity across occasions, the magnitudes and directions of exchange between clusters are described using occasion-specific weights. A recent alternative has been proposed by \cite{okada2025analysis}, who extended ACLUSKEW (Asymmetric Cluster analysis based on SKEW-symmetry) by \cite{okada2015asymmetric} from one-mode two-way asymmetric proximities to two-mode three-way asymmetric proximities.


Two-mode three-way data are referred to as matrix unconditional data when elements from any two different matrices can be meaningfully compared; otherwise, they are referred to as matrix conditional data.

Therefore, given the absence of existing methods for finding archetypal profiles for three-way asymmetric proximity data and the scarcity of clustering  and  multidimensional scaling methods (which are generally not scalable) for this kind of data, our research aims are as follows. The first aim is to extend the h-plot-based methodology proposed by \cite{Epi2013} to three-way proximity data, both for conditional and unconditional framework, and symmetric and asymmetric proximities. This extension offers the following novel advantages: intuitive interpretability through a unified Euclidean representation; an explicit, eigenvector-based analytical solution free from local minima; scale invariance under linear transformations; computational efficiency for large matrices (scalability); and a straightforward goodness-of-fit evaluation. The second aim is to use its results to obtain archetypal profiles for three-way proximity data; and the third aim is to cluster three-way asymmetric proximities. In Sec. \ref{background}, h-plot and ADA are reviewed. Sec. \ref{methodology} introduces our proposal for multidimensional scaling and finding archetypal profiles and clusters of three-way proximities. Our proposal is compared with several methodologies in Sec. \ref{compa}. A real application in a financial problem is analyzed in Sec. \ref{fina}. Finally, conclusions are provided in Sec. \ref{conclu}. The proofs are given in Appendix~\ref{sec:proofs}. 









\section{Background}
\label{background}

\subsection{H-plot for one-mode two-way data}


H-plots were introduced by \cite{CorstenGabriel} as a graphical method for visualizing relationships among multiple variables in a data matrix $\boldsymbol{X}$. Let $\boldsymbol{X}_c$ denote the centered data matrix, with $n$ objects and $m$ variables, and let $\boldsymbol{S}=\frac{1}{n-1}\boldsymbol{X}_c'\boldsymbol{X}_c$ 
be its variance-covariance matrix, where $'$ denotes transposition. Since $\boldsymbol{S}$ is positive semidefinite, its eigenvalues are nonnegative. Let $\lambda_1\geq\lambda_2\geq\cdots\geq\lambda_m\geq0$ denote the eigenvalues of $\boldsymbol{S}$, with corresponding unit eigenvectors $\boldsymbol{q}_1,\ldots,\boldsymbol{q}_m$. Let $\boldsymbol{Q}$ be the orthogonal matrix whose columns are these eigenvectors and let $\boldsymbol{\Lambda}=\operatorname{diag}(\lambda_1,\ldots,\lambda_m)$. The full h-plot is represented by the matrix $\boldsymbol{H}=\boldsymbol{Q}\boldsymbol{\Lambda}^{1/2}$, whose $i$-th row, denoted by $\boldsymbol{h}_i$, gives the coordinates of variable $i$. The two-dimensional h-plot is obtained by retaining the first two columns of $\boldsymbol{H}$, namely, $\boldsymbol{H}_2=
\left(
\sqrt{\lambda_1}\boldsymbol{q}_1,\,
\sqrt{\lambda_2}\boldsymbol{q}_2
\right).
$ In particular, the coordinates of variable $i$ in the two-dimensional h-plot are $
\boldsymbol{h}_i^{(2)}
=
\left(
\sqrt{\lambda_1}q_{i1},
\sqrt{\lambda_2}q_{i2}
\right).
$

The relationship between distances in the full h-plot and differences between variables follows directly from its matrix representation. Indeed, $
\boldsymbol{H}\boldsymbol{H}'
=
\boldsymbol{Q}\boldsymbol{\Lambda}\boldsymbol{Q}'
=
\boldsymbol{S}.
$ 
Consequently, for any two variables $i$ and $j$, $
\|\boldsymbol{h}_i-\boldsymbol{h}_j\|^2
=
S_{ii}+S_{jj}-2S_{ij}.
$ Using $\boldsymbol{S}=(n-1)^{-1}\boldsymbol{X}_c'\boldsymbol{X}_c$, we obtain

\begin{equation} \label{hplotdis} 
\|\boldsymbol{h}_i-\boldsymbol{h}_j\|^2
=
\frac{1}{n-1}
\left\|
\boldsymbol{X}_{c,i}-\boldsymbol{X}_{c,j}
\right\|^2 =\frac{1}{n-1}
\left\|
\boldsymbol{X}_i-\boldsymbol{X}_j
-
\left(\overline{X}_i-\overline{X}_j\right)\boldsymbol{1}
\right\|^2,
\end{equation}
where $\boldsymbol{X}_i$ denotes the $i$-th column of $\boldsymbol{X}$,
$\overline{X}_i$ its sample mean, and $\boldsymbol{1}$ the $n$-dimensional
vector of ones.

Hence, $
\|\boldsymbol{h}_i-\boldsymbol{h}_j\|
=
s\left(\boldsymbol{X}_{c,i}-\boldsymbol{X}_{c,j}\right),
$ where $s(\boldsymbol{X}_{c,i}-\boldsymbol{X}_{c,j})$ denotes the sample standard deviation of the differences between the centered observations of variables $i$ and $j$ ($\boldsymbol{X}_{c,i}$ denotes the $i$-th column of $\boldsymbol{X}_c$, containing the centered observations of variable $i$). Thus, distances in the full h-plot exactly represent the variability of pairwise differences between variables.

The two-dimensional h-plot $\boldsymbol{H}_2$ provides a low-dimensional approximation to this full representation. When the first two eigenvalues account for a large proportion of the total variance, $\boldsymbol{H}_2$ preserves most of the information contained in $\boldsymbol{H}$ while providing a convenient two-dimensional visualization of the relationships among variables.

Note that, although principal component analysis (PCA) and the h-plot approaches rely on the eigendecomposition of the variance-covariance matrix, their objectives differ. In PCA, the $n$ objects are represented in the component space through the scores, obtained by multiplying the centered data matrix by the matrix of eigenvectors. In contrast, the h-plot represents the $m$ variables through the eigenvectors scaled by the square roots of their corresponding eigenvalues.

To assess the quality of the two-dimensional h-plot representation, \cite{CorstenGabriel} proposed the following goodness-of-fit measure, with values close to 1 indicating a better representation:

\begin{equation}
\label{ajuste}
\frac{\lambda_1^2 + \lambda_2^2}{\sum_j \lambda_j^2}.
\end{equation}

This measure can be straightforwardly generalized to h-plots of any dimension by considering the corresponding subset of eigenvalues.

\subsubsection{H-plot for multidimensional scaling}
Let us see how h-plot can be used for multidimensional scaling. 
Let $\boldsymbol{\Delta}$ be an $n \times n$ dissimilarity matrix with elements $\delta_{ij}$ representing the dissimilarity from object $i$ to object $j$. When $\boldsymbol{\Delta}$ is asymmetric, we define
\[
\boldsymbol{D} = \left[ \boldsymbol{\Delta} \middle| \boldsymbol{\Delta}' \right],
\]
an $n \times 2n$ matrix formed by concatenating $\boldsymbol{\Delta}$ and its transpose $\boldsymbol{\Delta}'$ column-wise.

When $\boldsymbol{D}$ is treated as a data matrix, the first $n$ columns correspond to variables of the form $d_{\cdot j}$, that is, the dissimilarities from all other objects to object $j$. The second block of $n$ columns represents variables of the form $d_{j \cdot}$, that is, the dissimilarities from object $j$ to all other objects.

After applying h-plotting, the matrix $\boldsymbol{H}_2$ contains $2n$ rows: the first $n$ rows approximate the $d_{\cdot j}$ variables, while rows $n + 1$ through $2n$ approximate the $d_{j \cdot}$ variables.

In the context of multidimensional scaling,
h-plotting is applied to the matrix $\boldsymbol{\Delta}$ when it is symmetric, but to $\boldsymbol{D}$ if $\boldsymbol{\Delta}$ is asymmetric.  

In the latter case, we reverse the concatenation and partition the matrix $\boldsymbol{H}_2$ into two individual blocks, each consisting of $n$ rows, and display them.  The first block corresponds with $d_{\cdot j}$ profiles while the second block corresponds with $d_{j \cdot}$ profiles. As regards interpretation, if two profiles, $d_{\cdot j}$ and $d_{\cdot i}$, are similar, they will be positioned close to each other in the plot; the same holds analogously for the $d_{j \cdot}$ and $d_{i \cdot}$ profiles. Moreover, both $d_{\cdot j}$ and $d_{i \cdot}$ profiles are represented within a single unified space, allowing for direct comparison between them. In the symmetrical case, $d_{\cdot j}$ = $d_{j \cdot}$, so we only need to represent one variable for object $j$; this is why h-plotting is applied to matrix $\boldsymbol{\Delta}$.

In the asymmetrical case, when a $d_{\cdot j}$ profile lies near a $d_{j \cdot}$ profile, it indicates a high degree of similarity between them. Consequently, the Euclidean distance between $d_{\cdot j}$ and $d_{j \cdot}$ profiles within this representation can be used to assess asymmetry among objects, i.e. objects with close $d_{\cdot j}$ and $d_{j \cdot}$ profiles are more symmetric, whereas greater separation reflects stronger asymmetry.

In essence, the purpose of the h-plot is not to preserve pairwise dissimilarities exactly, but rather to maintain the relationships among the dissimilarity variables, that is, the dissimilarity profiles themselves. This makes h-plotting particularly useful for analyzing non-metric dissimilarities, where an exact Euclidean representation is not feasible due to the inherently non-Euclidean nature of the proximities.

If our goal is cluster or pattern detection, expanding or contracting the configuration may be more informative \citep{Seber}. Therefore, we can replace the actual dissimilarity values with their corresponding ranks instead \citep{borg2005modern,okada2024applied}. Similarly, other transformations (such as raising the dissimilarities to a power) can also be applied \citep{Podani}.


\subsubsection{Toy example}

For illustrative purposes, we consider the dissimilarities between four journals: an applied statistics one (AP), a journal of a specific statistical field (SF), a journal about statistical surveys (SU), and a theoretical statistical journal (TH), shown in Table \ref{revis}. This dissimilarity matrix is asymmetric and nonreflexive, since it is based on the number of citations between journals. Note that a small dissimilarity between journals $i$ and $j$ ($d_{ij}$) indicates that journal $i$ cites journal $j$ extensively. Cross-journal citations exhibit clear asymmetry: journal $i$ may not cite journal $j$ to the same extent that $j$ cites $i$. The triangle inequality may also fail, and self-citation indicates non-reflexivity.


\begin{table}[t]
\centering
\begin{tabular}{lcccc}
\hline
Citing/Cited
    & AP & SF & SU & TH \\ \hline
AP & 1  & 10 & 5  & 10 \\
SF & 4  & 2  & 5  & 10 \\
SU & 4  & 9  & 3  & 6  \\
TH & 10 & 10 & 7  & 1  \\ 
\hline
\end{tabular}
\caption{Dissimilarities between journals.}
\label{revis}
\end{table}

\begin{figure}[t]
\centering
\includegraphics[width=0.6\textwidth]{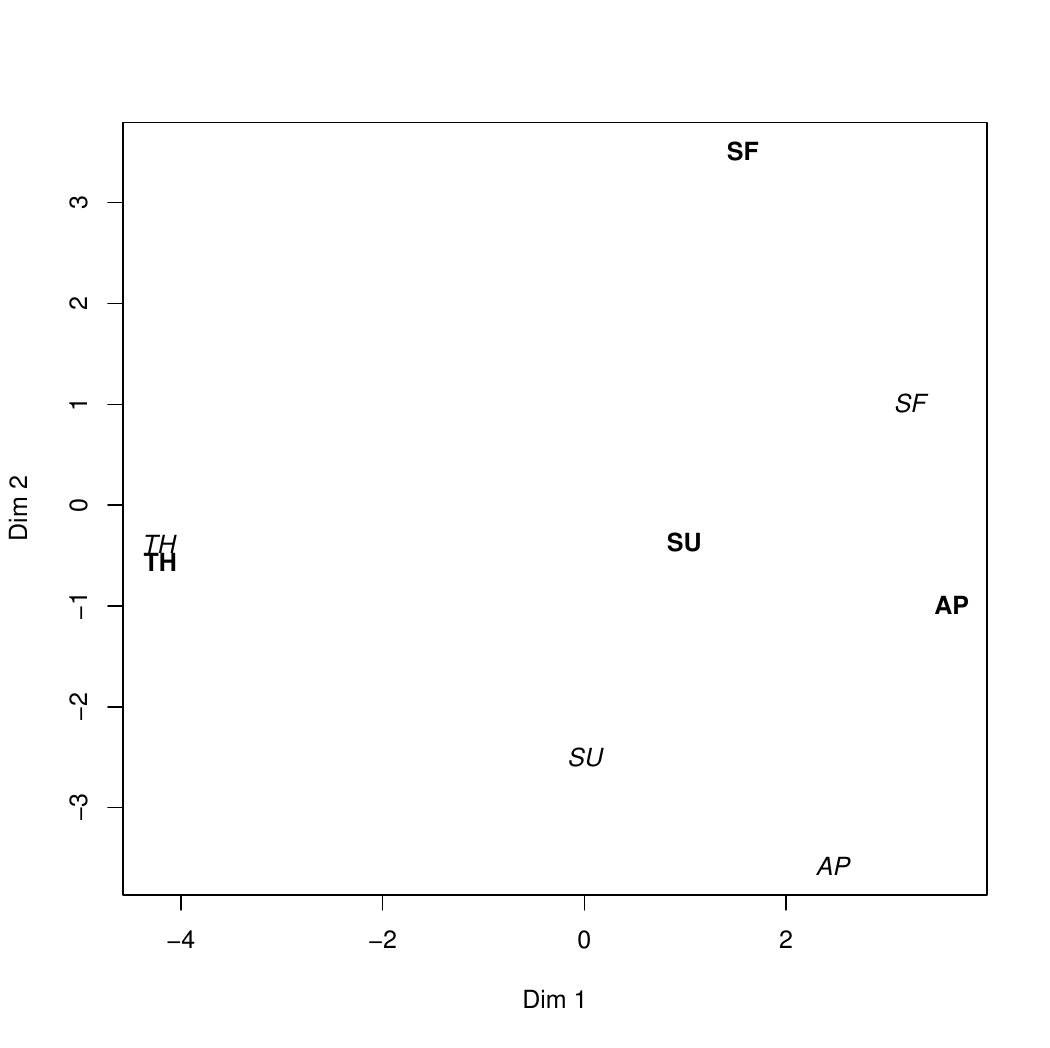}
\caption{H-plot projection of one-mode two-way journal citation data. Bold font represents the dissimilarity variables of cited ($d_{\cdot j}$), while italic font denotes citing ($d_{j \cdot}$).}\label{hplotrevis}
\end{figure}

Fig. \ref{hplotrevis} displayed the h-plot projection of Table \ref{revis}. The goodness of fit for the first dimension is 80.17\%, while it is 99.85\% for the two dimensions. The more applied journals are situated on the right part of the plot, while the theoretical journal is on the left-hand side. To assess the asymmetry for each journal, we compute the Euclidean distance between the h-plot representation of each journal as citing and cited. The most symmetrical journal (the journals that cite are similar to the journals that cited it) is TH, the theoretical journal; while the most asymmetrical journal is SF, whose cites come from different journals than those that the SF journal cites. 

Although the most citing and cited journal for SF is SF (the dissimilarity is 2), 
the nearest profile of SF citing profile is the cited AP profile, i.e., the journals that SF cites are more similar to the journals that cite AP. This is because the h-plot does not aim to reproduce pairwise dissimilarities directly, but rather to represent each object’s relationships with all others.

\subsubsection{Advantages of h-plots for multidimensional scaling}

H-plots offer several notable benefits:
\begin{enumerate}
    \item Intuitive interpretability: Profiles are directly comparable since all profiles are displayed within a single, unified Euclidean representation.
    
\item Explicit solution: They provide a direct analytical solution based on eigenvectors, avoiding the problem of local minima that often arises in iterative optimization methods. Furthermore, the resulting configuration remains consistent even when the roles of objects and individuals are interchanged.

\item Scale invariance: h-plots are invariant under linear transformations of the dissimilarity scale, meaning that changes in scale do not affect the overall visual configuration \citep{Epi2013}.

\item Computational efficiency: They are computationally efficient and capable of handling very large matrices, making them a scalable approach \citep{schwarz2019scalable}.

\item Simple goodness-of-fit evaluation: The quality of the representation can be easily quantified using the clear goodness-of-fit measure given by \eqref{ajuste}.

\item Robustness: When outliers are present, robust variants of h-plots can be computed, as proposed by \cite{daigle1992robust}.

\end{enumerate}

These advantages of the h-plot will be preserved in its extension to
two-mode three-way data.

\subsection{Partitioning around medoids (PAM)}

The $k$-medoids algorithm identifies $k$ representative observations within each cluster, referred to as medoids \citep{Kaufman90}, from the data set such that the total within-cluster dissimilarity is minimized, i.e., medoids correspond to actual data points from the original data set. It is defined for symmetrical dissimilarity matrices.

For a given $k$, $k$-medoids searches $k$ observations of the data,
${i_1}^*$, ..., ${i_k}^*$ such that
\begin{equation}
\{{i_1}^*, ..., {i_k}^*\} =
\operatorname*{arg\,min}_{\boldsymbol{Y}, {i_1}, ..., {i_k}}  \sum_{i \in \boldsymbol{Y}} \inf_{1 \leq j \leq k} \delta_{i i_j},
\end{equation}
where $\boldsymbol{Y}$ ranges on subsets of observations indexed from $1$ to $n$.  It can be computed with the function $\texttt{pam}$ from the \textsf{R} \citep{R} package \texttt{cluster} \citep{cluster}. This function also returns the silhouette information. The number of clusters for which the silhouette coefficient is maximum can be considered as the best choice of $k$ \citep{Kaufman90}.  If the average silhouette width exceeds 0.7, the clustering is regarded as strong; values near 0.5 indicate a reasonable solution, while those near 0.25 are generally viewed as weak.

\subsection{Archetypoid Analysis}

ADA is founded on the premise that data can be expressed as convex combinations of a small set of archetypal patterns drawn directly from the data set itself (see \cite{alcacer2025survey} for a comprehensive overview). In statistics, archetypal profiles carry the same intuitive meaning as in everyday language, i.e., they represent extreme or pure types. Archetypal patterns are more readily interpretable than central or average points, as people tend to understand data more intuitively through contrasting or opposing examples \citep{Thurau12}.

Let
$\boldsymbol{X} = \begin{pmatrix}
\boldsymbol{x}_1 \\
\vdots \\
\boldsymbol{x}_n
\end{pmatrix}$ denote an $n \times m$ data matrix containing $n$ observations and $m$ variables.
In ADA, three matrices are introduced to approximate the underlying data structure through \textit{mixtures} (i.e. convex linear combinations) of representative observations:

\begin{enumerate}
\item Archetypoids: The $k$ archetypoids, denoted by $\boldsymbol{z}_j$, form the rows of a $k \times m$ matrix $\boldsymbol{Z}$. Each $\boldsymbol{z}_j$, for $j=1,\ldots ,k$, corresponds to one specific case from the original data matrix $\boldsymbol{X}$.

\item Mixture coefficients: The $n \times k$ matrix $\boldsymbol{\alpha} = (\alpha_{ij})$ contains the coefficients used to approximate each observation $\boldsymbol{x}_i$ as a mixture of the archetypoids:
\begin{equation}
\label{eq:alpha}
\boldsymbol{x}_i\approx \hat{\boldsymbol{x}}_i = \sum_{j=1}^k \alpha_{ij} \boldsymbol{z}_j,
\end{equation}
where $\alpha_{ij} \geq 0$ and $\sum_{j=1}^k \alpha_{ij} = 1$ for $i = 1, \ldots, n$.

\item Selection matrix: The $k \times n$ matrix $\boldsymbol{\beta} = (\beta_{jl})$ is composed of binary coefficients that identify which observations are chosen as archetypoids. Each archetypoid $\boldsymbol{z}_j$ is defined as
\begin{equation}
\label{eq:beta}
\boldsymbol{z}_j = \sum_{l=1}^n \beta_{jl} \boldsymbol{x}_l,
\end{equation}
where $\beta_{jl} \in \{0, 1\}$ and $ \sum_{l=1}^n \beta_{jl} = 1$  for $j = 1, \ldots, k$, so it is guaranteed that each archetypoid corresponds to an actual case in $\boldsymbol{X}$.
\end{enumerate}

From \eqref{eq:alpha} and \eqref{eq:beta}, the Residual Sum of Squares ($\operatorname{RSS}$) is given by
\begin{equation} \label{RSSar}
\operatorname{RSS} = \sum_{i=1}^n \left( \boldsymbol{x}_i - \hat{\boldsymbol{x}}_i \right)^2
= \sum_{i=1}^n \left( \boldsymbol{x}_i - \sum_{j=1}^k \alpha_{ij} \sum_{l=1}^n \beta_{jl} \boldsymbol{x}_l \right)^2.
\end{equation}
The goal is to solve the following mixed-integer optimization problem, taking into account expression \eqref{RSSar}:
\begin{equation}
\label{eq:minrss}
\def\arraystretch{1.2}
\begin{array}[t]{rll}
\min \limits_{\boldsymbol{\alpha} ,\boldsymbol{\beta}} & \operatorname{RSS}\left( \boldsymbol{\alpha},\boldsymbol{\beta}\right) \\
\textrm{s.t.} & \displaystyle \sum_{j=1}^k \alpha_{ij} = 1,& \quad i=1,\ldots ,n, \quad \text{Convexity}\\
& \displaystyle \sum_{l=1}^n \beta_{jl} = 1,& \quad j=1,\ldots ,k, \quad \text{Binary assignment} \\
& \alpha_{ij}\geq 0, \, \beta_{jl} \in \{0, 1\}.&
\end{array}
\end{equation}

Program \eqref{eq:minrss} is solved using the two-phase algorithm introduced by \cite{Vinue15}. In the first phase, known as BUILD, an initial solution is generated by selecting a preliminary set of candidate archetypoids. The second phase, termed SWAP, iteratively refines this set by exchanging selected archetypoids with unselected observations whenever such replacements lead to a reduction in the $\operatorname{RSS}$. Our analysis employs the \textsf{R} implementation developed by \cite{EpiIbSi17}.

To decide the proper number of archetypoids, we adopt the elbow criterion, as used in prior studies such as \cite{cutler1994archetypal, Eugster2009, Vinue15}. This method entails plotting the $\operatorname{RSS}$ against the number of archetypoids and identifying the point where the rate of decrease in $\operatorname{RSS}$ markedly diminishes, commonly referred to as the \textit{``elbow''}.

\subsection{H-plot + ADA for one-mode two-way data}
\label{hplotADA}

For symmetrical relationships, ADA can be applied to the h-plot projections, since they are in an Euclidean space. Note that ADA assumes the data lie in a Euclidean space because the method relies on squared Euclidean distances and convex combinations. As archetypoids are sample points, the representative points are univocally identified. The ADA method has been successfully combined with h-plot projection in several previous studies. For instance, \cite{epifanio2023archetypal} applied it to the analysis of foot shapes, while \cite{math9070771} used it in a shoe size recommendation system. \cite{chapterAkinori26} also applied it to analyze cross-citation patterns in academic journals. Furthermore, \cite{doi:10.1080/00031305.2018.1545700} employed it across various data sets containing missing data. Additionally, \cite{Vinue15} applied the methodology to non-metric but symmetric dissimilarities derived from 3D binary images of the trunks of Spanish women for apparel design. In the same work, \cite{Vinue15} conducted a simulation study using non-metric yet symmetric dissimilarities to evaluate which MDS technique most effectively recovered archetypoids after projection. The h-plot method was compared with six other MDS techniques: Classical (Metric) MDS, Kruskal’s Non-metric MDS, Sammon’s Nonlinear Mapping, Isomap \citep{Tenenbaum}, Locally Linear Embedding \citep{RoweisLLE}, and Diffusion Map \citep{Coifman}. H-plot projections demonstrated superior performance in recovering the original archetypoids.

In the case of asymmetric relationships, \cite{vinue2017archetypoid} proposed the application of ADA to two-dimensional h-plot projections as well. It is important to note that each object $j$ has two associated profiles: one corresponding to $d_{j \cdot}$ and another approximating $d_{\cdot j}$. Both profiles are represented within the same configuration by the two-dimensional vectors $\boldsymbol{h}_j$ and $\boldsymbol{h}_{j+n}$, for $j = 1, \ldots, n$. Consequently, ADA is applied to an $n \times 4$ matrix $\boldsymbol{X}$, constructed by combining the representations from both blocks of h-plot profiles. Specifically, the $j$-th row of $\boldsymbol{X}$ consists of the concatenated vectors $\boldsymbol{h}_j$ and $\boldsymbol{h}_{j+n}$.

\section{Methodology}\label{methodology}
 Assume that $\boldsymbol{\Delta}_l$ $(l = 1, \ldots, L)$ denotes a square dissimilarity matrix, where each element $\delta_{ijl}$ corresponds to the pairwise dissimilarity from object $i$ to object $j$ $(i, j = 1, \ldots, n)$ observed at occasion or condition $l$. If $\delta_{ijl} = \delta_{jil}$, the dissimilarity matrix is said to be symmetric; otherwise, it is considered asymmetric. 
 \subsection{Multidimensional scaling for unconditional two-mode three-way data} \label{unconditional}

In this section, we consider unconditional two-mode three-way dissimilarity data, where the entries are directly comparable across occasions. The construction of the h-plot representation is based on treating the dissimilarity profiles corresponding to different occasions as variables in a common data matrix. As as consequence, for the symmetric case, we define
\[
\boldsymbol{D}_{\textrm{U}} = \left[ \boldsymbol{\Delta}_1 \middle| ... | \boldsymbol{\Delta}_L \right],
\] as an $n \times Ln$ matrix formed by concatenating $\boldsymbol{\Delta}_l$ matrices column-wise. While, for the asymmetric case, we define
\[
\boldsymbol{D}_{\textrm{U}} = \left[ \boldsymbol{\Delta}_1 \, \middle| \, \boldsymbol{\Delta}_1' \, \middle| \, \dots \, \middle| \, \boldsymbol{\Delta}_L \, \middle| \, \boldsymbol{\Delta}_L' \right],
\]
as an $n \times 2Ln$ matrix formed by concatenating the matrices $\boldsymbol{\Delta}_l$ and their transposes $\boldsymbol{\Delta}_l'$ column-wise.

\begin{proposition}
\label{prop:un}
Let $\boldsymbol{d}_a$ and $\boldsymbol{d}_b$ be any two dissimilarity
profiles represented as columns of $\boldsymbol{D}_{\textrm{U}}$, corresponding to
occasions $l$ and $s$, respectively. The squared Euclidean distance
between their coordinates in the full h-plot is
\[
\left\|
\boldsymbol{h}_a-\boldsymbol{h}_b
\right\|^2
=
\frac{1}{n-1}
\left\|
\boldsymbol{d}_a-\boldsymbol{d}_b
-
\left(\overline{d}_a-\overline{d}_b\right)\boldsymbol{1}
\right\|^2.
\]
Consequently, the h-plot allows direct comparisons between dissimilarity
profiles belonging to different occasions. In the symmetric case, the
profiles are of the form $\boldsymbol{d}_{\cdot jl}$, whereas in the
asymmetric case they may be of the form $\boldsymbol{d}_{\cdot jl}$ or
$\boldsymbol{d}_{j\cdot l}$.
\end{proposition}

For the symmetric case, $\boldsymbol{D}_{\textrm{U}}$ is h-plotted in two
dimensions, yielding the matrix $\boldsymbol{H}_2$ with $Ln$ rows and
two columns. Next, $\boldsymbol{H}_2$ is divided into $L$ blocks of
$n$ rows and rearranged into a matrix $\boldsymbol{Y}$ consisting of
$n$ rows and $2L$ columns, i.e., $L$ matrices of size $n\times 2$
stacked columnwise. Each of the $L$ blocks is visualized in a
two-dimensional plot, with distinct colors indicating the different
conditions.

For the asymmetric case, $\boldsymbol{D}_{\textrm{U}}$ is h-plotted in two
dimensions, yielding the matrix $\boldsymbol{H}_2$ with $2Ln$ rows and
two columns. We reverse the concatenation and partition
$\boldsymbol{H}_2$ into individual blocks, each consisting of $n$
rows. Each of the $2L$ blocks is visualized in a two-dimensional plot,
with $L$ distinct colors indicating the different conditions, and with
two different font styles: bold and italic for the
$d_{\cdot jl}$ and $d_{j\cdot l}$ profiles, respectively. The resulting
matrix, denoted by $\boldsymbol{Y}$, consists of $n$ rows and $4L$
columns. 

Although the method has been presented using a
two-dimensional projection for simplicity, it can be extended to
higher dimensions to capture additional information from
$\boldsymbol{D}_{\textrm{U}}$.

\subsubsection{Illustrative example}

\begin{figure}[t]
\centering
\includegraphics[width=0.8\textwidth]{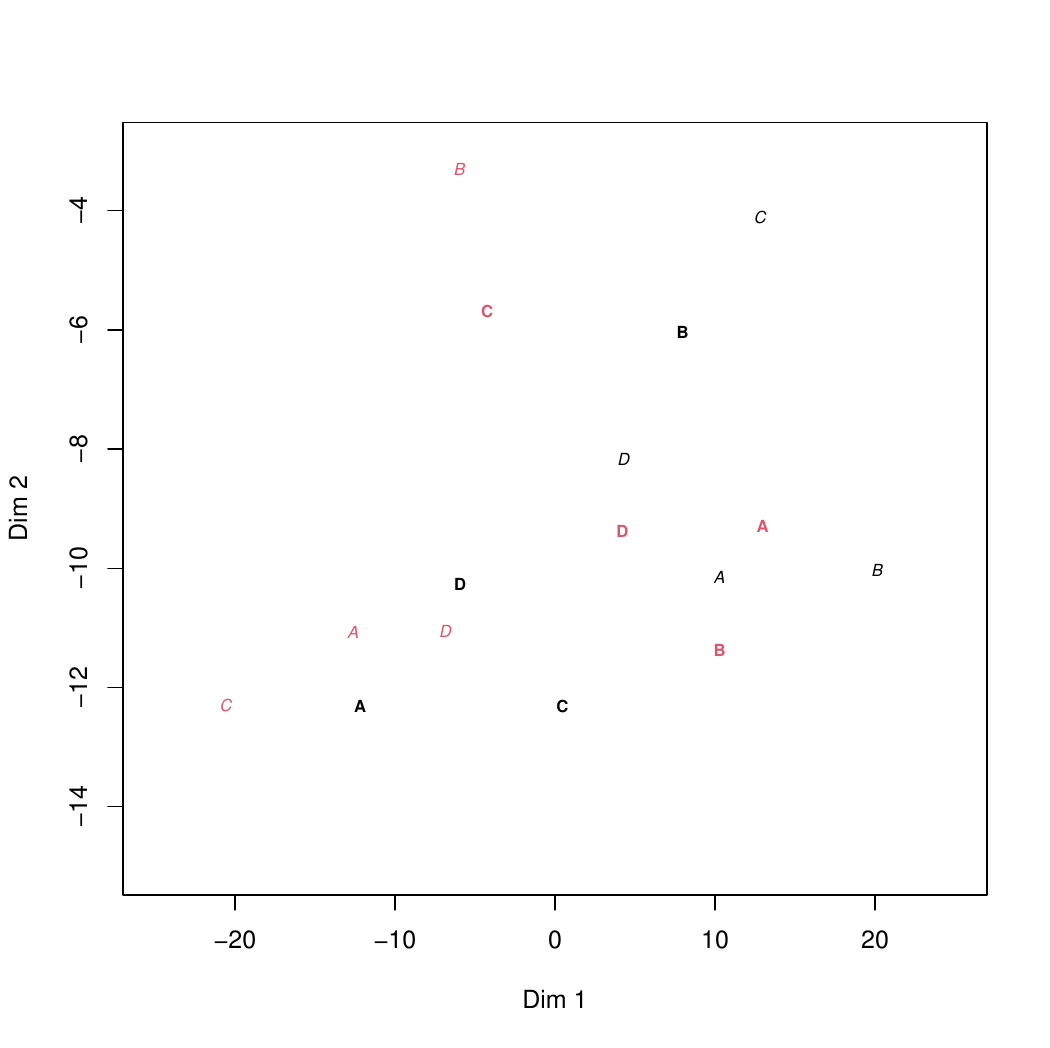}
\caption{H-plot projection of unconditional two-mode three-way artificial example. The first condition is represented in black, and the second in red. See the text for the meaning of the bold and italic fonts.}\label{fig1}
\end{figure}

Let us consider  in Table \ref{ilu} a three-way array representing pairwise exchanges of messages among $n = 4$ people in $L = 2$ occasions. Note that proximities are asymmetric, and the reflexive property is not fulfilled. Similarities are converted into dissimilarities by subtracting 50, i.e., the maximum value in the matrices. Considering that entries are comparable in both occasions, $\boldsymbol{D}$ is arranged as a $4 \times 16$ matrix in Table \ref{iluun}. Fig. \ref{fig1} displayed the h-plot projection. 

\begin{table}[t]
\centering
\begin{tabular}{c|cccc|cccc}
 & \multicolumn{4}{c|}{Occasion 1} & \multicolumn{4}{c}{Occasion 2} \\
\hline
   & A & B & C & D & A & B & C & D \\
\hline
A & 50 & 25 & 50 & 25 & 50 & 50 & 19 & 23 \\
B & 50 &  0 & 50 & 25 & 18 & 25 & 10 & 10 \\
C & 20 & 20 & 50 & 10 & 50 & 50 &  0 & 20 \\
D & 20 & 10 & 20 &  0 & 27 & 22 &  5 &  0 \\
\end{tabular}
\caption{Artificial three-way proximity data.}\label{ilu}
\end{table}

\begin{table}[t]
\centering
\begin{tabular}{c|cccc|cccc|cccc|cccc}
   & \multicolumn{4}{c|}{Received 1} 
   & \multicolumn{4}{c|}{Sending 1} 
   & \multicolumn{4}{c|}{Received 2} 
   & \multicolumn{4}{c}{Sending 2} \\
\hline
   & A & B & C & D & A & B & C & D & A & B & C & D & A & B & C & D \\
\hline
A &  0 & 25 &  0 & 25 &  0 &  0 & 30 & 30 &  0 &  0 & 31 & 27 &  0 & 32 &  0 & 23 \\
B &  0 & 50 &  0 & 25 & 25 & 50 & 30 & 40 & 32 & 25 & 40 & 40 &  0 & 25 &  0 & 28 \\
C & 30 & 30 &  0 & 40 &  0 &  0 &  0 & 30 &  0 &  0 & 50 & 30 & 31 & 40 & 50 & 45 \\
D & 30 & 40 & 30 & 50 & 25 & 25 & 40 & 50 & 23 & 28 & 45 & 50 & 27 & 40 & 30 & 50 \\
\end{tabular}
\caption{ Matrix $\boldsymbol{D}_{\textrm{U}}$ for the unconditional two-mode three-way artificial example. ``Received'' means $d_{\cdot j}$ profiles, while ``Sending'' denotes $d_{j \cdot}$ profiles, and numbers indicate the occasions 1 and 2.}
\label{iluun}
\end{table}

On the one hand, dimension 1 is related to interactions with individuals B and C. The further to the right, the less interaction with B and the more with C; while the further to the left, the more interaction with B and the less with C. On the other hand, dimension 2 measures interactions with A: the more interactions, the lower the position, and the fewer interactions, the higher the position. 

The most similar profiles, those that are nearest in h-plot configuration, are received by D in occasion 2 and sent by D in occasion 1. 

As regards asymmetry, the most symmetric profile is D in occasion 1, while the most asymmetrical is A in occasion 2, i.e., the received and sending profiles for D in occasion 1 are the most similar, but they are the least similar for A in occasion 2.

\subsection{Multidimensional scaling for conditional two-mode three-way data} \label{conditional}


In this section, we consider conditional two-mode three-way dissimilarity
data, where entries are comparable within each specific condition or
occasion (e.g., within a single participant's ratings), but not across
different conditions or occasions. Since the conditions are not
comparable, they cannot be treated as different dissimilarity variables
as in Section~\ref{unconditional}. Instead, the conditions are treated
as distinct observations of the dissimilarity variables. Since the
covariances between the dissimilarity variables are computed to
construct the h-plot, comparisons within the same row are valid.

For the symmetric case, we define
\[
\boldsymbol{D}_{\textrm{C}} =
\begin{bmatrix}
\boldsymbol{\Delta}_1 \\
\vdots \\
\boldsymbol{\Delta}_L
\end{bmatrix},
\]
as an $Ln \times n$ matrix formed by concatenating the matrices
$\boldsymbol{\Delta}_l$ row-wise.

For the asymmetric case, we define
\[
\boldsymbol{D}_{\textrm{C}} =
\begin{bmatrix}
\boldsymbol{\Delta}_1 \mid \boldsymbol{\Delta}_1' \\
\vdots \\
\boldsymbol{\Delta}_L \mid \boldsymbol{\Delta}_L'
\end{bmatrix},
\]
as an $Ln \times 2n$ matrix formed by concatenating the matrices
$\boldsymbol{\Delta}_l$ and their transposes $\boldsymbol{\Delta}_l'$
row-wise.

The distinction between the two constructions, $\boldsymbol{D}_{\textrm{U}}$ and $\boldsymbol{D}_{\textrm{C}}$,  lies in the role of the
occasion index. Under the unconditional construction, the occasion
index is part of the definition of the dissimilarity profile, so that
profiles from different occasions belong to the same variable space.
In contrast, under the conditional construction, the occasion index
refers to repeated observations of the same profile variables and is
therefore not part of the variable definition.

The conditional construction is appropriate when the occasions are
regarded as repeated observations of the same dissimilarity variables
and their covariance structure is comparable across occasions. When
location and/or scale differences across occasions are considered
nuisance effects, we recommend standardizing the dissimilarities within
each occasion before stacking. Under this transformation, the h-plot
is based on the covariance structure of the standardized dissimilarities
across occasions, rather than on differences in their marginal levels
or scales.

\begin{proposition} \label{prop:cond} 
For any two dissimilarity profiles corresponding to columns $j$ and
$k$ of $\boldsymbol{D}_{\textrm{C}}$, the squared Euclidean distance between their coordinates in the
full h-plot is
\[
\left\|
\boldsymbol{h}_j-\boldsymbol{h}_k
\right\|^2
=
\frac{1}{Ln-1}
\sum_{l=1}^{L}\sum_{i=1}^{n}
\left[
\left(\delta_{ijl}-\delta_{ikl}\right)
-
\overline{\left(\delta_{\cdot jl}-\delta_{\cdot kl}\right)}
\right]^2
\]
in the symmetric case. In the asymmetric case, the same result holds
for any pair of dissimilarity profiles represented by columns of
$\boldsymbol{D}_{\textrm{C}}$, whether they correspond to columns of
$\boldsymbol{\Delta}_l$ or $\boldsymbol{\Delta}_l'$. Consequently,
the h-plot compares dissimilarity profiles within each occasion $l$,
without introducing comparisons between different occasions.
\end{proposition}    

For the symmetric case, $\boldsymbol{D}_{\textrm{C}}$ is h-plotted in two
dimensions, yielding the matrix $\boldsymbol{H}_2$ with $n$ rows and
two columns. This h-plot projection, denoted by $\boldsymbol{Y}$,
consists of $n$ rows and $2$ columns.

For the asymmetric case, $\boldsymbol{D}_{\textrm{C}}$ is h-plotted in two
dimensions, yielding the matrix $\boldsymbol{H}_2$ with $2n$ rows and
two columns. We reverse the concatenation and partition
$\boldsymbol{H}_2$ into two individual blocks, each consisting of $n$
rows. Each of the two blocks is visualized in a two-dimensional plot,
with two different font styles: bold and italic for
$\boldsymbol{d}_{\cdot j}$ and $\boldsymbol{d}_{j\cdot}$ profiles,
respectively. This new matrix, denoted by $\boldsymbol{Y}$, consists
of $n$ rows and $4$ columns. 

As before, although the approach has been
presented in terms of a two-dimensional projection for simplicity, it
can be generalized to higher dimensions in order to capture additional
information from $\boldsymbol{D}_{\textrm{C}}$.

\subsubsection{Illustrative example}
\label{iluc}

\begin{figure}[t]
\centering
\includegraphics[width=0.8\textwidth]{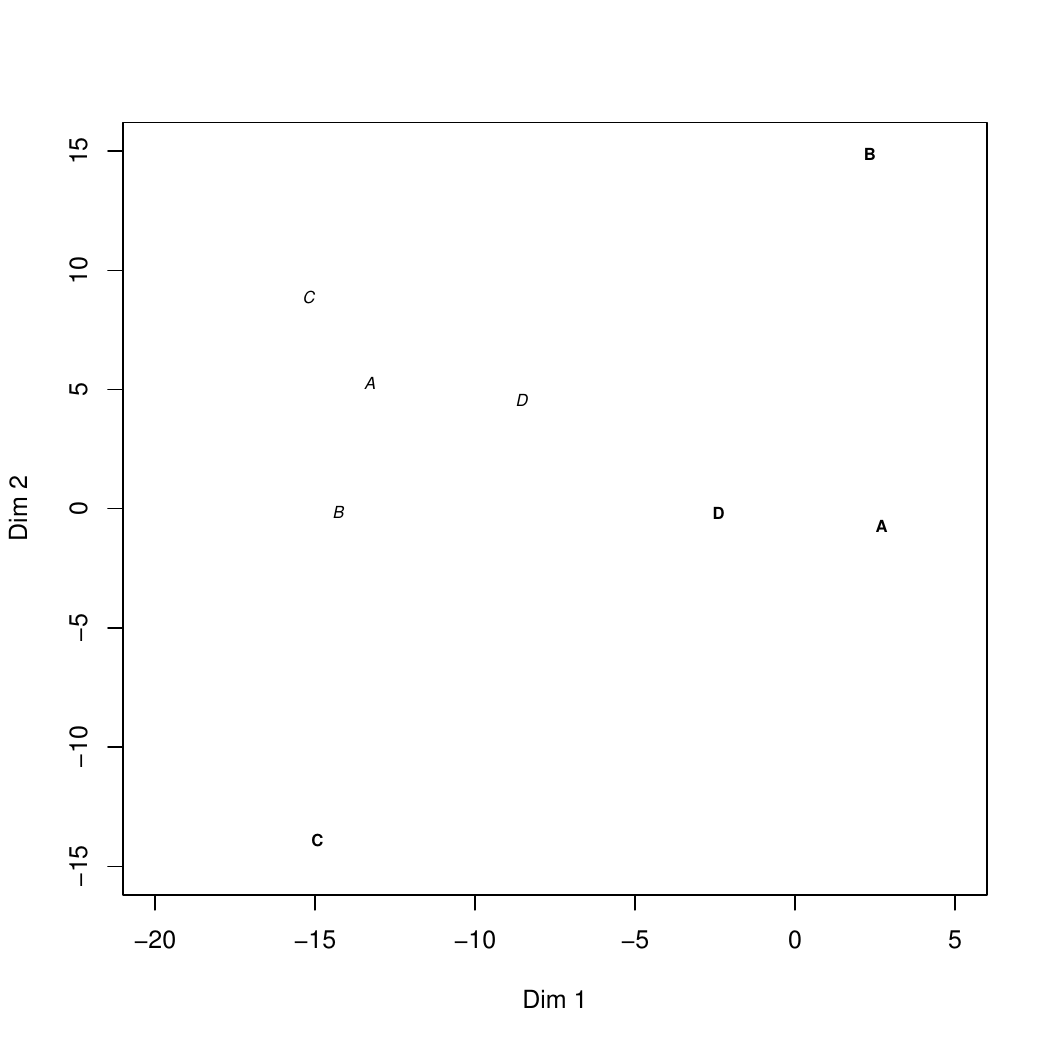}
\caption{H-plot projection of conditional two-mode three-way artificial example. See the text for the meaning of the bold and italic fonts.}\label{fig2}
\end{figure}

Data in Table \ref{ilu} is considered. Again, similarities are converted into dissimilarities by subtracting 50 to all entries in the matrix. Now, we consider that the entries in both occasions are not comparable, therefore  $\boldsymbol{D}_{\textrm{C}}$ is arranged as an $8 \times 8$ matrix as shown in Table \ref{ilucon}. Its h-plot is displayed in Fig. \ref{fig2}. 

\begin{table}[t]
\centering
\begin{tabular}{c|cccc|cccc}
 & \multicolumn{4}{c|}{\textbf{Received}} & \multicolumn{4}{c}{\textbf{Sending}} \\
\hline
\textbf{Occasion 1} & A & B & C & D & A & B & C & D \\
\hline
A &  0 & 25 &  0 & 25 &  0 &  0 & 30 & 30 \\
B &  0 & 50 &  0 & 25 & 25 & 50 & 30 & 40 \\
C & 30 & 30 &  0 & 40 &  0 &  0 &  0 & 30 \\
D & 30 & 40 & 30 & 50 & 25 & 25 & 40 & 50 \\
\hline
\textbf{Occasion 2} & A & B & C & D & A & B & C & D \\
\hline
A &  0 &  0 & 31 & 27 &  0 & 32 &  0 & 23 \\
B & 32 & 25 & 40 & 40 &  0 & 25 &  0 & 28 \\
C &  0 &  0 & 50 & 30 & 31 & 40 & 50 & 45 \\
D & 23 & 28 & 45 & 50 & 27 & 40 & 30 & 50 \\
\end{tabular}
\caption{ Matrix $\boldsymbol{D}_{\textrm{C}}$ for the conditional two-mode three-way artificial example.}\label{ilucon}
\end{table}

The first dimension is related to interactions with individual C. The further to the right, the  less interaction with C in occasion 1 and the more with C in occasion 2;
while the further to the left, the more interaction with C in occasion 1 and the less in occasion 2. On the other hand, dimension 2 measures interactions with A and B: the fewer interactions in occasion 1 and the more interactions in occasion 2, the higher the position, and conversely,  the more interactions with A and B in occasion 1 and the fewer in occasion 2, the lower the position.

The most similar profiles, those that are nearest in h-plot configuration, are sent by A and sent by C. 

Regarding asymmetry, the most symmetric profile is D, while the most asymmetrical is C. Specifically, the received and sending profiles for D are the most similar, whereas they are the least similar for C.

\subsection{Archetypal profiles for two-mode three-way data}
\label{ADA23}


In an analogous way to that in Sect. \ref{hplotADA}, ADA is applied to
the matrix $\boldsymbol{Y}$, which differs for unconditional and
conditional two-mode three-way data. The proposition \ref{prop:euclidean}
establishes the Euclidean nature of the h-plot representation and,
consequently, the validity of applying ADA to $\boldsymbol{Y}$. By
contrast, ADA cannot be applied directly to dissimilarities, regardless of their
arrangement. Since ADA relies on Euclidean distances and convex combinations of
observations, the dissimilarity profiles must first be represented in
a Euclidean space where these operations are well defined.

\begin{remark}
\label{prop:euclidean}
The h-plot represents the dissimilarity profiles by coordinates in a
Euclidean space obtained from the eigendecomposition of the
variance-covariance matrix of the corresponding data matrix. In both the unconditional and conditional frameworks, the h-plot
coordinates provide a Euclidean representation of the corresponding
dissimilarity profiles. Consequently, Euclidean distances and convex
combinations are well defined in the resulting space, and ADA can be
validly applied to $\boldsymbol{Y}$.
\end{remark}

\subsection{Clustering for two-mode three-way data}
\label{cluster24}

Similarly, as explained in Sect. \ref{ADA23}, we can apply a clustering method  to matrix  $\boldsymbol{Y}$. We consider PAM for obtaining medoids, which are representative points, like the alter ego of archetypoids for clustering.





\section{Comparison with other methodologies}
\label{compa}
In this section, the proposed methodologies are compared with several existing alternative approaches using three real data sets previously analyzed in the literature and a simulation study. The evaluation criteria depend on the type of method and the information available for comparison. 

For multidimensional scaling methods, there is no standard goodness-of-fit measure specifically designed for asymmetric 3-way 2-mode proximity data that would provide a common basis for comparing all the considered approaches. Furthermore, the representations obtained from the different models cannot be directly compared using a common coefficient, as they rely on different interpretations of the proximity data. Our approach represents the dissimilarities in a Euclidean space, whereas alternative approaches, such as radius-based models, rely on a different representation and interpretation. Consequently, following \citet[Sec. 20.7]{borg2005modern}, graphical representations are used as the primary means of assessing and comparing the resulting configurations.

For archetypal profiles, no existing alternative method provides directly comparable archetypal profiles, and therefore a quantitative comparison with another method is not applicable. 

For clustering approaches, in contrast, the simulation study provides a known ground-truth partition, allowing the recovery of the true clustering structure to be quantitatively assessed using the Adjusted Rand Index (ARI) \citep{hubert1985comparing}.

\subsection{Illustrative example for multidimensional scaling}

\begin{figure}[t]
\centering
\includegraphics[width=0.8\textwidth]{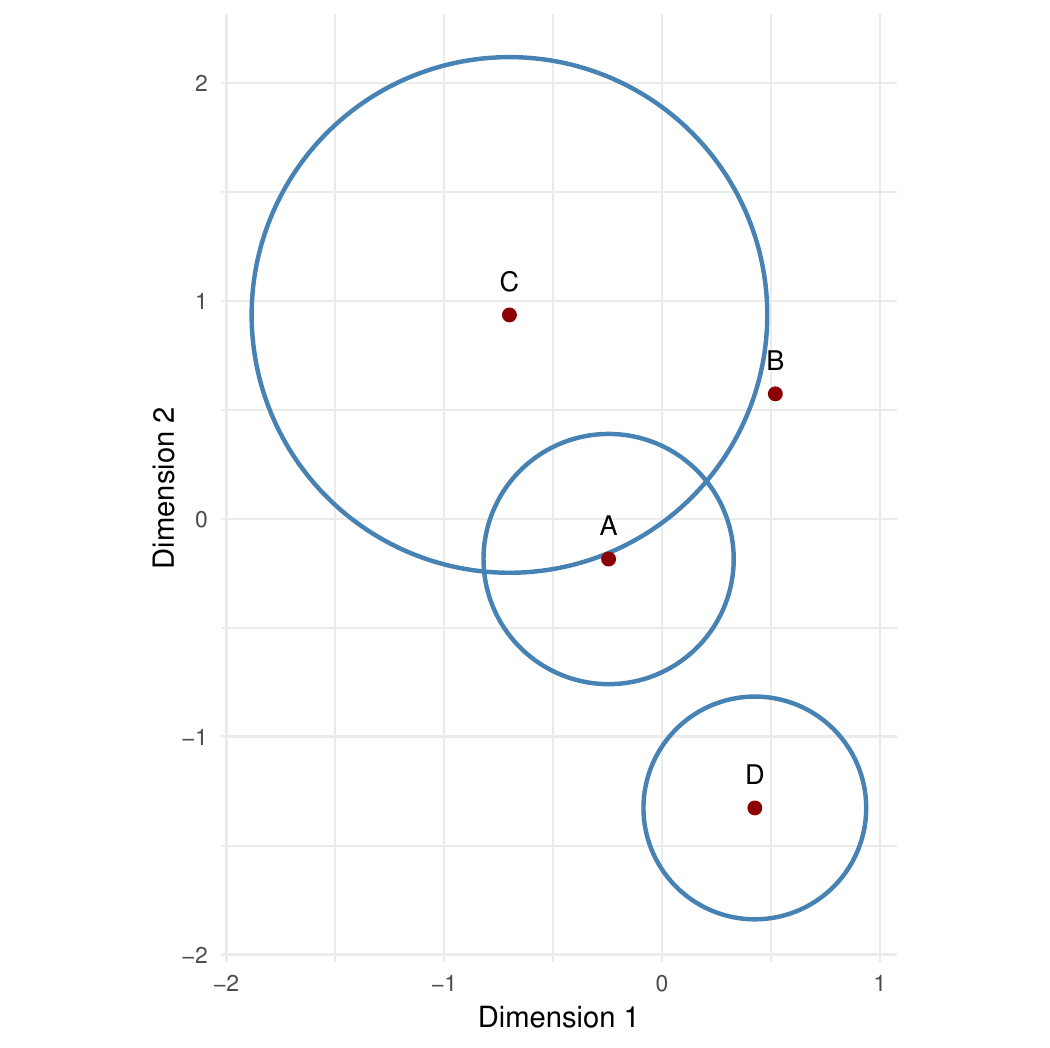}
\caption{Two-dimensional common object configuration of illustrative example in Sec. \ref{iluc} by radius-distance model.}\label{ilucradius}
\end{figure}

The data in Table \ref{ilu} are analyzed using the two-mode three-way radius-distance model developed by \cite{okada1997asymmetric}, which is matrix conditional. Figure \ref{ilucradius} shows the results. The common object configuration shown in this figure reveals similar symmetric relationships among the objects, while the radii provide information about their asymmetry. The most asymmetric person is C, which is consistent with the h-plot results shown in Figure \ref{fig2}. For the most symmetric case, however, the two representations differ, as B has a radius of zero in the radius-distance model. This difference illustrates the complementary information provided by the two approaches. While the radius-distance model summarizes the asymmetry of each object through a single radius, the h-plot provides separate representations of the sent and received profiles, allowing differences between these two types of profiles to be examined directly.

\subsection{Asian Nation Data  for multidimensional scaling} 

The following example was thoroughly analyzed by \cite{okada2024applied}. The data set comprises two $10 \times 10$ asymmetric similarity matrices (i.e., $10 \times 10 \times 2$ two-mode, three-way similarity data) representing relationships among 10 Asian countries. In each matrix, the element at position ($j$, $i$) of the $l$-th matrix denotes the mean rating (provided by respondents in nation $j$) of the influence exerted by nation $i$ on nation $j$ during year $l$. A larger value in the ($j$, $i$) cell indicates that respondents in nation $j$ perceived the influence of nation $i$ on their own nation more positively or favorably in that year. \cite{okada2024applied} analyzed this data set using the two-mode three-way radius-distance model developed by \cite{okada1997asymmetric}, which is matrix conditional, and INDSCAL \citep{carroll1970analysis}, which assumes symmetry and therefore cannot model asymmetric relationships. 

\begin{figure}[t]
\centering
\includegraphics[width=0.8\textwidth]{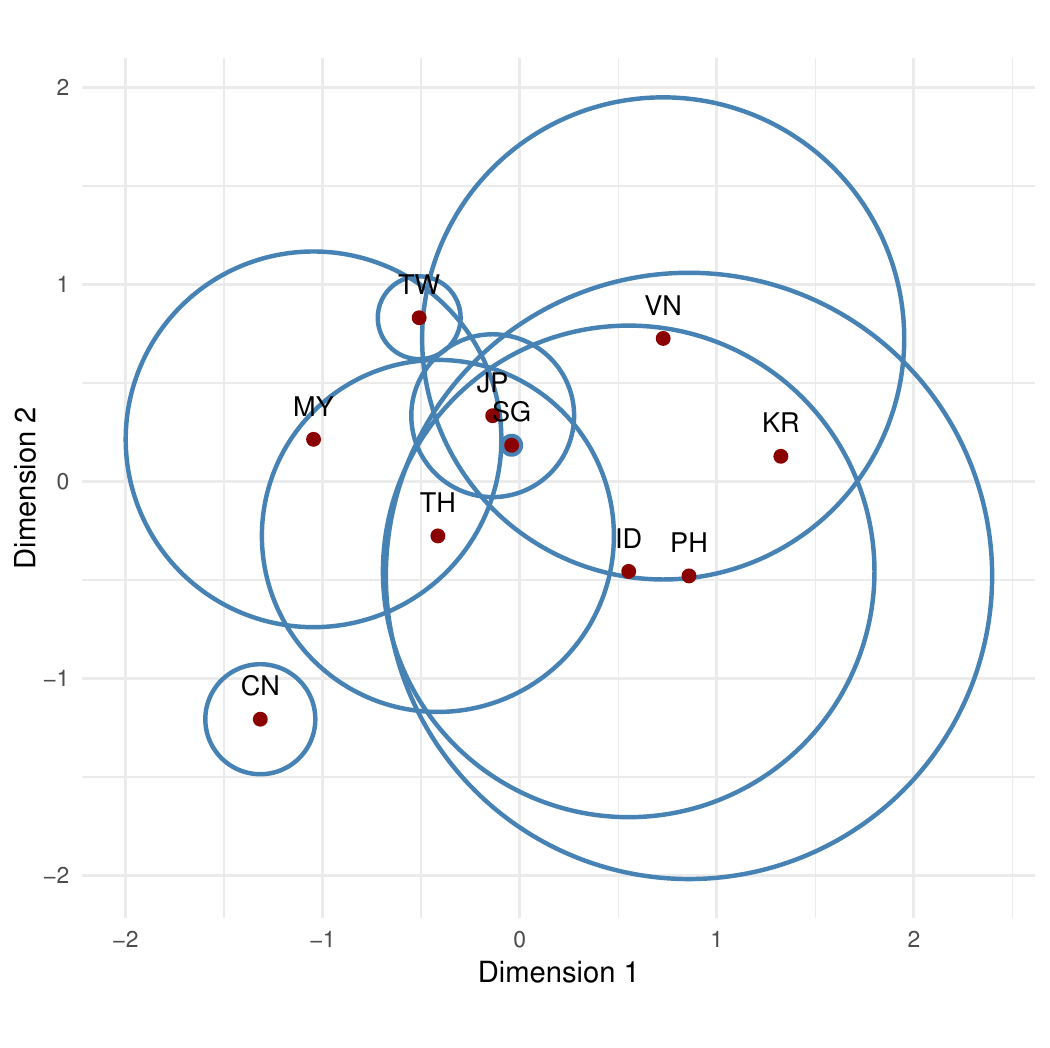}
\caption{Two-dimensional common object configuration of nation data set by radius-distance model.}\label{nations24}
\end{figure}

Fig. \ref{nations24} shows the results with the radius-distance model.  Korea has the smallest radius, equal to zero, indicating that respondents in Korea consistently evaluate the influence of other nations on Korea as more negative or less favorable than Korea’s influence on them. In contrast, the Philippines has the largest radius, suggesting that respondents there tend to rate the influence of other nations on the Philippines as more positive or favorable than the reverse. The correlation coefficients of GDP and each dimension and radius are -0.5, -0.61, and -0.34, respectively.

\begin{figure}[t]
\centering
\includegraphics[width=0.8\textwidth]{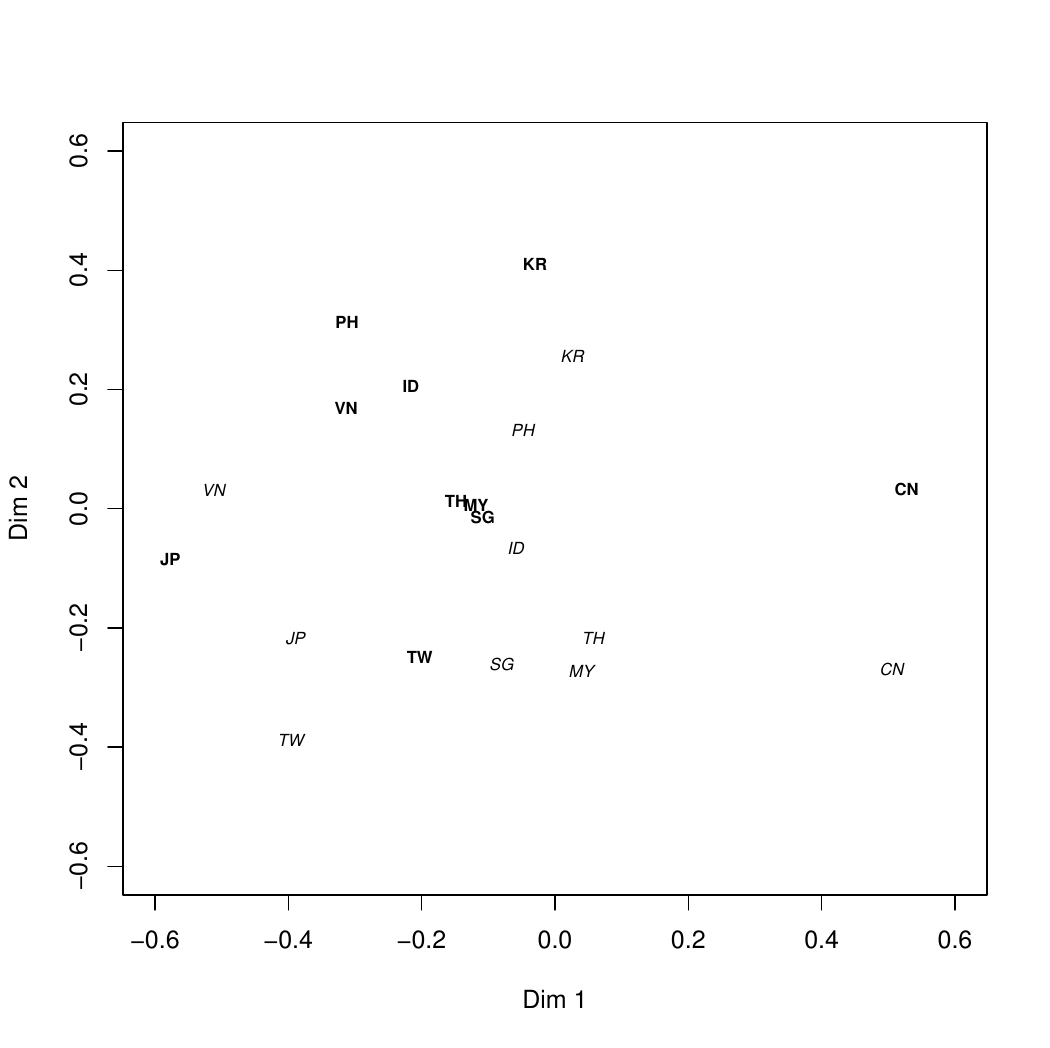}
\caption{H-plot projection of nation data set. Bold font represents influencing ($d_{i \cdot}$) profiles, while italic font denotes influenced ($d_{\cdot i}$) profiles.}\label{nationshplot}
\end{figure}

Although our model can work with conditional and unconditional matrices, we will show only the h-plot projection for the conditional case, in order to make a fair comparison with the results produced by the radius-distance model. Fig. \ref{nationshplot} shows the h-plot for this data set. The goodness of fit for two dimensions is 78\%, while the first dimension explains 61\%. The correlation between GDP and the first h-plot dimension is 0.64 for the influencing profiles, and 0.58 for the influenced ones.  The most symmetric country is Korea, while the most asymmetric are Malaysia and the Philippines (both tied). The most similar profiles are the influencing from Malaysia, Singapore, and Thailand. 

Results that emerge from Figs. \ref{nations24} and \ref{nationshplot} have coincidences, although  there are also some differences worth mentioning. For example, Japan profiles are somewhat separated from the rest of the countries in the h-plot, something that does not happen with the radius-distance model. 

The fact that Japan is clearly separated from the other countries in the h-plot, but not in the radius-distance representation, illustrates that the two approaches capture different aspects of asymmetric relationships. The separation of Japan in the h-plot may reflect a distinctive pattern of perceived international influence that is not captured by the radius-distance representation. In particular, the h-plot represents separately the influencing and influenced profiles of each country, and can therefore distinguish countries according to the overall pattern of their relationships, rather than summarizing their asymmetry through a single radius. Thus, Japan may exhibit a distinctive configuration of perceived influence despite not being characterized by an especially large degree of asymmetry.



\subsection{Simulated data for clustering} 
We considered a simulation study with $n=30$ objects belonging to $k=3$ equally sized clusters, with 10 objects per cluster. The true cluster membership was kept fixed across two occasions, resulting in a three-way asymmetric dissimilarity array of dimension $30\times30\times2$. Let $g_i\in\{1,2,3\}$ denote the true cluster membership of object $i$. For each occasion $l=1,2$, the off-diagonal dissimilarities were generated according to
\[
\delta_{ijl}
=
\mu_{g_i g_j}^{(l)}+\varepsilon_{ijl},
\qquad i\neq j,
\]
where $\varepsilon_{ijl}\overset{\mathrm{iid}}{\sim}N(0,\sigma^2)$, with $\sigma=0.5$, and the diagonal elements were set to zero. After generating the dissimilarities, negative values were truncated at zero. The expected dissimilarity between objects $i$ and $j$ was therefore determined by their cluster memberships and by the occasion-specific mean dissimilarity matrix $M^{(l)}$.

We considered two simulation scenarios that differed in the degree of asymmetry of the underlying dissimilarity structure. In the first scenario, the mean dissimilarity matrices for the two occasions were
\[
M^{(1)}=
\begin{pmatrix}
1.00&1.30&1.60\\
0.80&1.00&0.90\\
0.70&1.35&1.00
\end{pmatrix},
\qquad
M^{(2)}=
\begin{pmatrix}
1.05&1.35&1.45\\
0.85&1.05&0.95\\
0.75&1.40&1.05
\end{pmatrix}.
\]
These matrices represent a pronounced asymmetric structure, with substantial differences between reciprocal cluster pairs. The underlying structure is similar across the two occasions, while the mean dissimilarities differ slightly between occasions.

In the second scenario, the mean dissimilarity matrices were
\[
M^{(1)}=
\begin{pmatrix}
1.00&1.30&1.35\\
1.20&1.00&1.20\\
1.30&1.35&1.00
\end{pmatrix},
\qquad
M^{(2)}=
\begin{pmatrix}
1.05&1.35&1.40\\
1.25&1.05&1.25\\
1.35&1.40&1.05
\end{pmatrix}.
\]
This second scenario also generates asymmetric dissimilarities, but the differences between reciprocal cluster pairs are smaller, representing a less pronounced degree of asymmetry. As in the first scenario, the same cluster structure is maintained across the two occasions, while the occasion-specific mean matrices allow for moderate changes in the dissimilarity levels.

For each scenario, $50$ independent data sets were generated. In every replication, the same true partition was used across both occasions, whereas the dissimilarities were independently perturbed by Gaussian noise. To assess the clustering performance, we compared our proposed methodology with the approach of \cite{bocci2024clustering}, considering its complete partition, which is associated with the symmetric component of the data, as opposed to the incomplete partition, which captures the corresponding imbalances. The clustering of the unconditional two-mode three-way data was obtained according to our procedure described in Sect.~\ref{cluster24}. Since the true cluster membership is known, the clustering solutions obtained by both methods were evaluated against the ground truth using the ARI. The ARI ranges from $-1$ to $1$, with values close to $1$ indicating a high agreement between the estimated and true partitions, whereas values close to $0$ indicate agreement comparable to that expected by chance. 

In the first scenario, characterized by a more pronounced asymmetric dissimilarity structure, the proposed h-plot methodology achieves a higher mean ARI than the approach of \cite{bocci2024clustering} ($0.684$ vs.~$0.615$), although with greater variability ($0.21$ vs.~$0.189$). In contrast, in the second scenario, where the degree of asymmetry is less pronounced, the approach by \cite{bocci2024clustering} performs better, with a mean ARI of $0.941$ compared with $0.824$ for h-plot + PAM, and also shows lower variability ($0.095$ vs.~$0.163$). Thus, the proposed h-plot methodology appears particularly competitive when the asymmetric structure is more pronounced. This pattern is consistent with the different aspects of the asymmetric structure captured by the two approaches. The complete partition in \cite{bocci2024clustering} primarily captures the average amounts of exchange, whereas the incomplete partition accounts for the corresponding imbalances. Consequently, the advantage of the proposed h-plot methodology in the scenario with more pronounced asymmetry suggests that explicitly representing the directional structure may be particularly beneficial in this setting.

Regarding computational cost, for only $n=30$ objects, the proposed h-plot methodology is essentially instantaneous ($8 \times 10^{-4}$ seconds per replication), whereas the method of \cite{bocci2024clustering}, using $10$ random initializations, requires approximately $27.03$ seconds in one replication. The computational time of the latter method also increases considerably as the number of objects grows. Thus, even for this relatively small problem size, the proposed methodology is substantially faster. All computations were performed on a computer equipped with an Intel\textsuperscript{\textregistered} Core\textsuperscript{\texttrademark} Ultra 9 275HX processor and 64 GB of DDR5 RAM, running Windows 11 and \textsf{R} version 4.5.1.

\subsection{Student Mobility Data for clustering}

This data set was subjected to detailed analysis by \cite{bocci2024clustering}. The data used in this study come from the OECD Education Statistics Database, which provides annual information on international student exchanges in tertiary education. The data set comprises five asymmetric matrices representing student mobility among 20 OECD founding countries from 2016 to 2020.  Each matrix records the number of students moving from an origin country to a host country for tertiary studies. For each year, the proportions of outgoing students were converted into pairwise dissimilarities by calculating 100 minus each proportion. The countries are the twenty founding
members of the OECD: Austria (AT), Belgium (BE), Canada (CA), Denmark (DK), France
(FR), Germany (DE), Greece (EL), Iceland (IS), Ireland (IE), Italy (IT), Luxembourg (LU), Netherlands (NL), Norway (NO), Portugal (PT), Spain (ES), Sweden (SE), Switzerland (CH), Turkey (TR), United Kingdom (UK), and United States of America (US). 

\cite{bocci2024clustering} computed 6 clusters, composed by: 
1) DE, US; 2) EL, IE, TR; 3) DK, IS, NO, SE; 4) BE, FR, LU, NL, PT, ES; 5) AT, CA, IT, CH; and 6) UK.

We have performed the clustering of the unconditional two-mode three-way data following the procedure described in Sect. \ref{cluster24}. According to the silhouette information, the optimal number of clusters is $k=4$, with a silhouette coefficient of $0.53$. Although $k=4$ is preferred according to this criterion, we also report the results for $k=6$ to facilitate comparison with \cite{bocci2024clustering}. The medoid of each cluster is listed first. For $k=4$, the clusters obtained are: 1) DE, AT; 2) FR, BE, EL, IT, LU, NL, PT, ES, CH, TR; 3) UK, CA, IE, US; 4) NO, DK, IS, SE. For $k=6$, the clusters obtained are: 1) DE, AT; 2) IT, BE, LU, NL, CH; 3) UK, CA, US; 4) NO, DK, IS, SE; 5) ES, EL, FR, PT, TR; 6) IE.

Results returned by our proposal seem more geographically coherent than those obtained with the methodology of \cite{bocci2024clustering}. This is noteworthy because geographical proximity, together with historical, cultural, linguistic, and educational ties, can be expected to play an important role in international student mobility. From this perspective, some of the groups identified by \cite{bocci2024clustering}, such as DE and US, CA together with IT and CH, or IE together with EL and TR, are less expected, as they combine countries that are geographically distant and, in some cases, have substantially different educational and mobility contexts. In contrast, our approach tends to identify groups of countries with greater geographical coherence, suggesting that the proposed representation may capture meaningful regional patterns in student mobility.

\begin{figure}[t]
\centering
\includegraphics[width=0.7\textwidth]{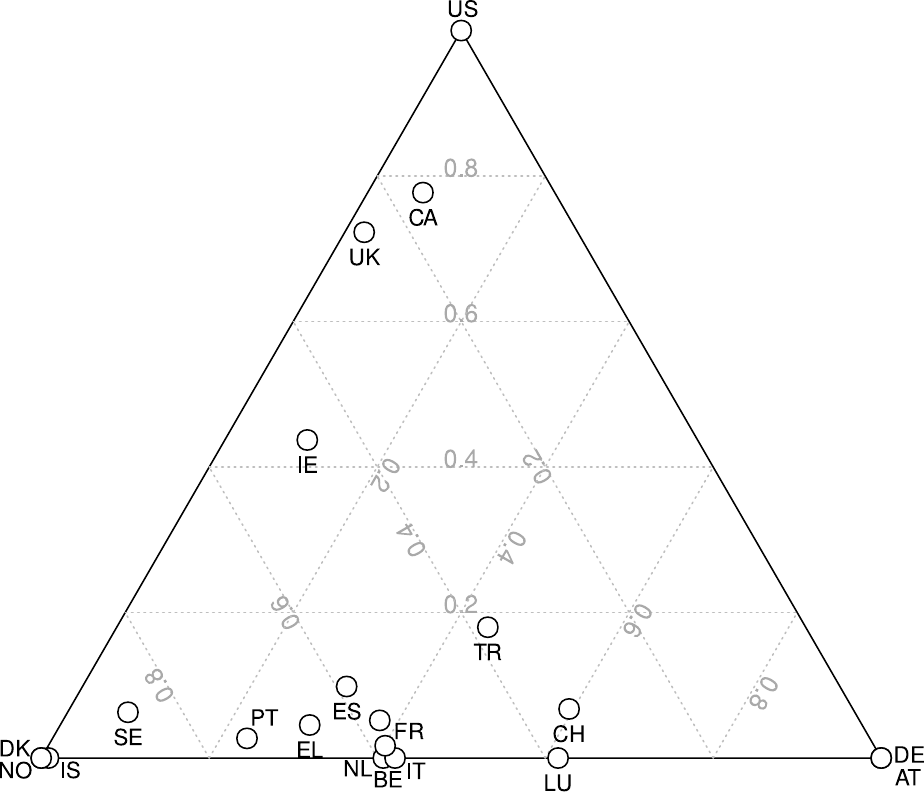}
\caption{Ternary plot of alpha values of ADA for student mobility data.}\label{ternarycountries}
\end{figure}

We have also estimated the archetypoids following the procedure proposed in Sect. \ref{ADA23}. According with the screeplot, the best $k$ is $3$. Alpha values are displayed in a ternary plot in Fig. \ref{ternarycountries}. The archetypoids are: NO, AT, and US. The rest of the countries are approximated by mixtures of these archetypoids. For example, BE is approximated by a mixture of 59\% NO and 41\% AT; UK is approximated by a mixture of 25\% NO, 2\% AT, and 72\% US. 

\subsection{Cultural Participation Data for clustering}

This data set was analyzed in detail  by  \cite{okada2025analysis}. The data set consists of $9 \times 9 \times 7$ two-mode, three-way asymmetric similarity matrices that describe cultural participation relationships. Each matrix represents how participation in one cultural activity relates to another across seven social groups. The nine cultural activities included are (we denote  the labels for each activity in parentheses): (a) Classical music performances and concerts (Classic), (b) Museum and art exhibitions (Museum),  (c) Hobby and cultural lessons (Lesson), (d) Travel abroad (Abroad), (e) Volunteer activities (Volunteer), 
(f) Public library use (Library), 
(g) Reading novels or history books (Novels), 
(h) Karaoke (Karaoke), (i) Reading tabloid papers and celebrity magazines (Tabloid). 

The two-mode three-way ACLUSKEW developed by   \cite{okada2025analysis} is matrix unconditional. They considered two clusters, whose dominant objects are Tabloid and Novels, which are shared across all social groups. The former represents lowbrow culture, while the latter reflects a middlebrow orientation. 

We have calculated the clustering of the unconditional two-mode three-way data following our procedure described in Sect. \ref{cluster24}. We consider two groups for each social group for comparison purposes. Table \ref{cluster25} displays ACLUSKEW's results, the dominance objects for ACLUSKEW, and the medoids for PAM are indicated in bold. 

\begin{table}[ht]
\centering
\scriptsize
\begin{tabular}{p{2cm}| p{2.5cm} p{2.5cm}|p{2.5cm} p{2.5cm}}
Social group                             & \multicolumn{2}{c|}{ACLUSKEW}                                                                                                                                    & \multicolumn{2}{c}{H-plot + PAM}        \\ \hline
                                           & Cluster 1                                                                            & Cluster 2                                            & Cluster 1 & Cluster 2 \\ \hline
Female junior high school graduates                  & \textbf{Tabloid}, Classic, Museum, Lessons, Abroad, Volunteer, Library, Karaoke & \textbf{Novels}                                               &      \textbf{Abroad}, Lessons, Volunteer               &    \textbf{Library}, Classic, Museum, Novels,    Karaoke, Tabloid              \\ \hline
Female high school graduates                         & \textbf{Tabloid}, Museum, Lessons, Abroad, Volunteer, Library, Karaoke                                            & \textbf{Novels},  Classic                                                       &           \textbf{Abroad}, Lessons, Volunteer        &  \textbf{Novels}, Classic, Museum, Library,    Karaoke, Tabloid                    \\ \hline
Female junior college or technical college graduates & \textbf{Tabloid}, Lessons, Abroad, Library, Karaoke                                                               & \textbf{Novels}, Classic, Museum, Volunteer                                    &          \textbf{Abroad}, Lessons, Volunteer         &           \textbf{Novels}, Classic, Museum, Library,    Karaoke, Tabloid        \\ \hline
Female college graduates                             & \textbf{Tabloid}                                                                                             & \textbf{Novels}, Classic, Museum, Lessons, Abroad, Volunteer, Library, Karaoke &     \textbf{Abroad}, Classic, Lessons, Volunteer              &    \textbf{Novels}, Museum, Library,    Karaoke, Tabloid              \\ \hline
Male junior high school graduates                    & \textbf{Tabloid}, Classic, Museum, Lessons, Abroad, Volunteer, Library, Karaoke                                   & \textbf{Novels}                                                            &       \textbf{Lessons}, Classic, Abroad, Volunteer            &   \textbf{Novels},  Museum, Library,    Karaoke, Tabloid               \\ \hline
Male high school graduates                           & \textbf{Tabloid}, Classic, Museum, Lessons, Abroad, Volunteer, Library, Karaoke                                   & \textbf{Novels}                                                             &           \textbf{Abroad}, Classic, Lessons, Volunteer        &  \textbf{Novels},  Museum, Library,    Karaoke, Tabloid               \\ \hline
Male junior college graduates or above               & \textbf{Tabloid}                                                                                             & \textbf{Novels}, Classic, Museum, Lessons, Abroad, Volunteer, Library, Karaoke &        \textbf{Abroad},  Classic, Lessons, Volunteer          &   \textbf{Library},   Museum, Novels,    Karaoke, Tabloid             \\ \hline
\end{tabular}
\caption{Clustering results for ACLUSKEW and h-plot+PAM for Cultural Participation Data, with two groups.} \label{cluster25}
\end{table}

The ACLUSKEW results yield single-member clusters for four of the seven social groups. By contrast, our proposal produces more balanced cluster sizes across all social groups. Moreover, the clustering structure obtained with the h-plot + PAM approach is more consistent across social groups, with similar groups of cultural activities being identified for different demographic categories.

We have also estimated the clusters for $k=2$ without distinguishing between social groups. In this case, the resulting clusters are (medoids are shown in bold): 1) \textbf{Abroad}, Lessons, Volunteer; and 2) \textbf{Novels}, Classic, Museum, Library, Karaoke, Tabloid. Nevertheless, according to the silhouette information, the optimal number of clusters is $k=3$, with a silhouette coefficient of $0.49$. The resulting clusters are: 1) \textbf{Abroad}, Lessons, Volunteer; 2) \textbf{Novels}, Museum, Library, Karaoke, Tabloid; and 3) \textbf{Classic}.

Interestingly, the two-cluster solution obtained with the h-plot + PAM approach provides a different perspective to the ACLUSKEW solution. While ACLUSKEW identifies Tabloid and Novels as the dominant objects of the two clusters, our approach identifies Abroad, Lessons, and Volunteer as a distinct cluster. The differences between the two solutions may reflect the distinct ways in which the two methods represent the underlying structure of the cultural participation data, rather than indicating that one solution is preferable to the other.

We have also estimated the archetypoids following the procedure proposed in Sect. \ref{ADA23}. According to the screeplot, the best $k$ is $3$. The archetypoids are: Volunteer, Tabloid, and Classic. If we assign each activity to the archetypoid with the highest $\alpha$ value, the groups obtained coincide with those provided by PAM.


\section{Application to financial products}
\label{fina}

The data set for this application corresponds to that used in \cite{Ferrer21}, consisting of daily closing prices of green bonds and equities, as well as some conventional asset classes that represent investment alternatives. Specifically, we consider the following equity indices from the NASDAQ OMX Green Economy family: Solar Energy (GRNSOLAR), Wind Energy (GRNWIND), Fuel Cell (GRNFUEL), Energy Efficiency (GRNENEF), Clean Transportation (GRNTRN), Green Buildings (GRNGB) and Pollution Prevention (GRNPOL). Moreover, we also consider the S\&P Green Bond Index (SPGBI) and some conventional asset classes: S\&P GSCI Gold (GOLD), Brent oil price (OIL) and MSCI World Index (MSCI). Finally, the global Treasury, investment-grade, and high-yield corporate bond markets are proxied by the Bloomberg Barclays Global Treasury Total Return Index (TREAS), the Bloomberg Barclays Aggregate Corporate Bond Index (IGBOND), and the Bloomberg Barclays Global High Yield Corporate Total Return Index (HYBOND). See \cite{Ferrer21} for a full description of these financial products.

Asymmetric dissimilarities ranging from $0$ to $1$ are computed using the oriented squared wavelet coherence introduced by \cite{Ferrer21}, which is based on the squared wavelet coherence and the wavelet phase-difference \citep{Torrence99}. These dissimilarities are defined in terms of causality, such that if object $i$ influences object $j$, then the dissimilarity from object $i$ to object $j$ is smaller than the dissimilarity from object $j$ to object $i$. Three sub-periods associated with some major financial, economic, and even health-related events are examined: the European sovereign debt crisis (13 October 2010 to 31 July 2012), the oil price collapse  (20 June 2014 to 28 February 2016), and the 2020 COVID-19 pandemic (1 January to 13 November 2020). 
We have considered a scale interval of more than $22$ days, which captures long-term interactions. The reason is that relationships between asset markets over longer time periods are less affected by short-lived phenomena, such as shifts in investor sentiment or psychological factors, than relationships over shorter time periods.

Therefore, our data are composed of a three-way data array consisting of three asymmetric dissimilarity matrices (each of which considers a different sub-period) of 14 observations. 

\subsection{Results}

\begin{figure}[t]
\centering
\includegraphics[width=1\textwidth]{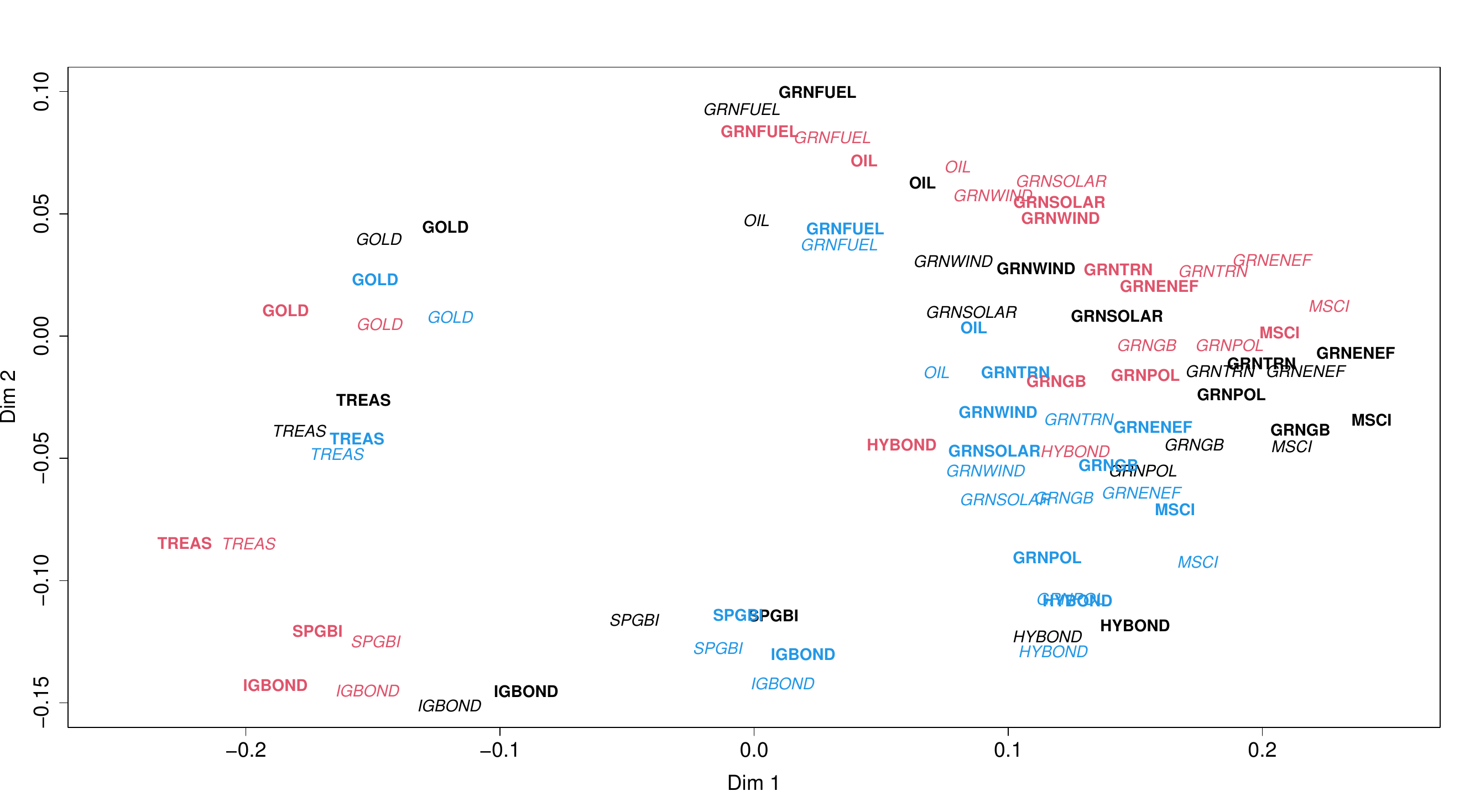}
\caption{H-plot projection of financial products data set. Bold font represents influencing ($d_{i \cdot}$) profiles, while italic font denotes influenced ($d_{\cdot i}$) profiles. The first sub-period is represented in black, the second in red, and the third in blue.}\label{long}
\end{figure}

We have computed the h-plot (see  Fig. \ref{long}) for the unconditional two-mode three-way data.  
The goodness of fit is $94\%$. The green stocks are clustered on the right part of the plot, together with OIL and HYBOND, in accordance with the clusters obtained by \cite{Ferrer21} for the long term. Furthermore, we can observe that these products are a bit shifted between each sub-period. On the one hand, the most asymmetrical profiles are OIL and HYBOND. Since, in most cases, the dissimilarities from OIL to the other assets are greater than the dissimilarities from the other assets to OIL (mainly in the second and third sub-periods), we can conclude that OIL influences others (particularly green stocks) more than it is influenced by them. In contrast, the dissimilarities from HYBOND to the other assets are less than the dissimilarities from the other assets to HYBOND, concluding that HYBOND is influenced more than it influences. On the other hand, the less asymmetrical profiles are TREAS and GRNFUEL. This suggests that these assets are not closely linked to the others in terms of cause and effect.

Archetypoids are calculated following the procedure proposed in Sect. \ref{ADA23}. According to the screeplot, the best $k$ is $4$, being the archetypoids: TREAS,  IGBOND,  MSCI, and GRNFUEL.  The remaining assets are estimated to be a mixture of these archetypoids. For example, GOLD is expressed as 70\% TREAS and 30\% GRNFUEL. Furthermore, the green stocks are mainly a mixture of GRNFUEL and MSCI, with GRNFUEL having a greater weighting. This mixture is not surprising, given that green equities represent a significant proportion of the overall equity market as measured by the MSCI index.


\section{Conclusions}
\label{conclu}

We have extended the h-plot methodology for multidimensional scaling to the analysis of three-way proximity data, encompassing both symmetric and asymmetric as well as conditional and unconditional frameworks. While traditional MDS methods focus primarily on symmetric data, the proposed approach effectively addresses the challenges posed by asymmetric and three-way proximities, which have remained largely unexplored.

The extended h-plot framework offers several notable and novel advantages over previous methods for two-mode
three-way asymmetric dissimilarities: intuitive interpretability within a unified Euclidean representation; an explicit, eigenvector-based analytical solution that avoids local minima; scale invariance under linear transformations; high computational efficiency for large data sets; and a simple, quantitative goodness-of-fit assessment.

Moreover, the proposed methodology facilitates the identification of archetypal profiles and clustering structures for three-way asymmetric proximities, enhancing interpretability and analytical flexibility. Comparative analyses with existing MDS and clustering approaches, supported by a financial application, demonstrate its effectiveness and complementary value, providing additional insights into asymmetric data. All data and code are made publicly available to ensure full reproducibility and transparency.

Several avenues for future research emerge from this work. On the one hand, some extensions to the methodology to other kinds of data could be
developed. One promising direction involves representing symbolic dissimilarities using h-plots. In many applications, exact dissimilarities may be unavailable; however, partial information such as their range \citep{Groenen} or their distribution represented as histograms \citep{groenen2006multidimensional} may be available. For instance, interval dissimilarities could be visualized through h-plots by adapting the principles proposed by \cite{d2021principal}. As regards missing values in the proximity matrices, several alternatives could be explored, as this problem has been considered in  PCA  \citep{gomez2024solving}. On the other hand, a promising line of future research lies in the application of the proposed methodology to practical problems across various fields, such as economics and finance or psychology. Finally,  a modification of biarchetypal analysis \citep{alcacer2024biarchetype} could be developed to identify archetypal profiles within proximity data.

\section*{Funding}
This work was partially supported by the Spanish Ministry of Science and Innovation PID2023-153128NB-I00, PID2022-141699NB-I00 and PID2020-118763GA-I00 and Generalitat Valenciana CIPROM/2023/66.

\section*{Declaration of competing interest}
The authors have declared that no competing interests exist.

\section*{Data and code availability}
For reproducibility, codes and data are available in \url{https://epifanio.uji.es/RESEARCH/asy2mode3way.zip}.

\appendix

\section{Proofs}
\label{sec:proofs}

\begin{proof}[Proof of Proposition \ref{prop:un}]
The result follows directly from the distance property of the h-plot
given in \eqref{hplotdis}, by taking $\boldsymbol{X}=\boldsymbol{D}_{\textrm{U}}$.
Thus, for any two columns $\boldsymbol{d}_a$ and $\boldsymbol{d}_b$ of
$\boldsymbol{D}_{\textrm{U}}$,
\[
\left\|
\boldsymbol{h}_a-\boldsymbol{h}_b
\right\|^2
=
\frac{1}{n-1}
\left\|
\boldsymbol{d}_{a,c}-\boldsymbol{d}_{b,c}
\right\|^2.
\]
In the symmetric case, taking
\[
\boldsymbol{d}_a=\boldsymbol{d}_{\cdot jl},
\qquad
\boldsymbol{d}_b=\boldsymbol{d}_{\cdot ks},
\]
gives
\[
\left\|
\boldsymbol{h}_{\cdot jl}-\boldsymbol{h}_{\cdot ks}
\right\|^2
=
\frac{1}{n-1}
\left\|
\boldsymbol{d}_{\cdot jl}
-
\boldsymbol{d}_{\cdot ks}
-
\left(
\overline{d}_{\cdot jl}
-
\overline{d}_{\cdot ks}
\right)\boldsymbol{1}
\right\|^2.
\]
Hence, when $l\neq s$, the h-plot directly compares dissimilarity
profiles belonging to different occasions. The same argument applies
in the asymmetric case to profiles represented by columns of
$\boldsymbol{\Delta}_l$ and $\boldsymbol{\Delta}_l'$.

\end{proof}

\begin{proof}[Proof of Proposition \ref{prop:cond}]
The result follows directly from \eqref{hplotdis}, applied with
$\boldsymbol{X}=\boldsymbol{D}_{\textrm{C}}$. In the symmetric case, the $j$-th
and $k$-th columns of $\boldsymbol{D}_{\textrm{C}}$, denoted by
$\boldsymbol{d}_{\cdot j}$ and $\boldsymbol{d}_{\cdot k}$, respectively,
are obtained by stacking the corresponding columns of
$\boldsymbol{\Delta}_1,\ldots,\boldsymbol{\Delta}_L$. Hence,
\[
\boldsymbol{d}_{\cdot j}
=
\begin{pmatrix}
\delta_{1j1}\\
\vdots\\
\delta_{nj1}\\
\vdots\\
\delta_{1jL}\\
\vdots\\
\delta_{njL}
\end{pmatrix},
\qquad
\boldsymbol{d}_{\cdot k}
=
\begin{pmatrix}
\delta_{1k1}\\
\vdots\\
\delta_{nk1}\\
\vdots\\
\delta_{1kL}\\
\vdots\\
\delta_{nkL}
\end{pmatrix}.
\]
Applying \eqref{hplotdis} gives
\[
\begin{aligned}
\left\|
\boldsymbol{h}_j-\boldsymbol{h}_k
\right\|^2
&=
\frac{1}{Ln-1}
\left\|
\boldsymbol{d}_{\cdot j,c}
-
\boldsymbol{d}_{\cdot k,c}
\right\|^2\\
&=
\frac{1}{Ln-1}
\sum_{l=1}^{L}\sum_{i=1}^{n}
\left[
\left(\delta_{ijl}-\delta_{ikl}\right)
-
\overline{\left(\delta_{\cdot jl}-\delta_{\cdot kl}\right)}
\right]^2.
\end{aligned}
\]

In the asymmetric case, the same argument applies to the columns
arising from $\boldsymbol{\Delta}_l$ and $\boldsymbol{\Delta}_l'$,
which represent, respectively, the profiles
$\boldsymbol{d}_{\cdot j}$ and $\boldsymbol{d}_{j\cdot}$. Thus, in
each difference, the dissimilarities correspond to the same occasion
$l$. Therefore, no comparisons between different occasions are
introduced.
\end{proof}

\bibliography{hplot}
\bibliographystyle{abbrvnat}

\end{document}